\documentclass[sigconf]{acmart}

\renewcommand\footnotetextcopyrightpermission[1]{} 

\AtBeginDocument{%
  }

\usepackage{multirow}
\usepackage{pifont}
\usepackage{subcaption} 
\usepackage{graphicx}   
\usepackage{enumitem}

\usepackage[skip=3pt]{caption}

\newcommand{\system}{\textsc{LLMigrate}}

\usepackage{titlesec}
\titlespacing*{\section}{0pt}{*0.9}{*0.9} 
\titlespacing*{\subsection}{0pt}{*0.8}{*0.8}
\titlespacing*{\subsubsection}{0pt}{*0.8}{*0.8}

\setlist[enumerate]{itemsep=0.3pt, topsep=0.5pt}
\setlist[itemize]{itemsep=0.3pt, topsep=0.5pt}

\makeatletter
\renewcommand{\paragraph}{%
  \@startsection{paragraph}{4}%
  {\z@}{0.2ex \@plus 0.2ex \@minus .2ex}{-1em}%
  {\normalfont\normalsize\bfseries}%
}
\makeatother

\begin{document}

\title{LLMigrate: Transforming ``Lazy'' Large Language Models into Efficient Source Code Migrators}

\author{Yuchen Liu}
\affiliation{%
  \institution{Peking University}
  \city{Beijing}
  \country{China}}

\author{Junhao Hu}
\affiliation{%
  \institution{Peking University}
  \city{Beijing}
  \country{China}}

\author{Yingdi Shan\textsuperscript{*}}
\affiliation{%
  \institution{Tsinghua University}
  \city{Beijing}
  \country{China}}
  
\author{Ge Li}
\affiliation{%
  \institution{Peking University\\ Zhongguancun Laboratory}
  \city{Beijing}
  \country{China}}

\author{Yanzhen Zou}
\affiliation{%
  \institution{Peking University\\ Zhongguancun Laboratory}
  \city{Beijing}
  \country{China}}

\author{Yihong Dong}
\affiliation{%
  \institution{Peking University}
  \city{Beijing}
  \country{China}}

\author{Tao Xie}
\affiliation{%
  \institution{Peking University}
  \city{Beijing}
  \country{China}}

\begin{abstract}

Rewriting C code in Rust provides stronger memory safety, yet migrating large codebases such as the 32-million-line Linux kernel remains challenging. While rule-based translators (e.g., C2Rust) provide accurate yet largely unsafe Rust programs, recent Large Language Model (LLM) approaches produce more idiomatic, safe Rust programs but frequently exhibit \emph{laziness}, omitting significant portions of the target code.

To address the issue, in this paper, we present \system, an LLM-based C-to-Rust translation tool that splits modules into discrete functions, translating them individually, and then reintegrating them. \system \ uses static analysis to retain necessary context, pairs GPT-4o (a state-of-the-art LLM) with compiler-driven translation and program-repair techniques for complex core functions, and leverages call-graph–guided translation to ensure consistent interfaces. Evaluations on three representative Linux kernel modules (math, sort, and ramfs) show that \system \  requires modifying less than 15\% of the target code, significantly outperforming a pure GPT-4o–based migration.

\end{abstract}

\ccsdesc[500]{Software and its engineering~Software creation and management}
\ccsdesc[500]{Computing methodologies~Artificial intelligence}

\keywords{Code Translation, Large Language Models}



\maketitle

\renewcommand{\thefootnote}{\fnsymbol{footnote}}
\footnotetext[1]{Corresponding author.}
\renewcommand{\thefootnote}

\pagestyle{plain} 
\section{Introduction}
\label{sec:intro}

In recent years, developers have increasingly transitioned applications originally written in C to the Rust programming language~\cite{DBLP:conf/usenix/LiGYWX24, DBLP:conf/esem/PanterE24}. This shift is motivated by Rust's robust memory safety guarantees and its performance, which rivals traditional languages such as C and C++. However, migrating complex systems faces significant challenges due to their large and continuously expanding codebases. For instance, the Linux kernel, one of the most critical software systems, had grown to 32.2 million lines of code across 74,300 files as of December 2021, with its size continuing to increase.



To accelerate migration and reduce manual effort, an automated C2Rust translation tool~\cite{c2rust} has been proposed but such tool typically translates C programs into correct but unsafe Rust programs~\cite{c2rust}, motivating us to design an effective tool for translating C code of large-scale systems to safe Rust for addressing two main challenges. First, achieving high accuracy is challenging. Accuracy and labor efficiency are two sides of the same coin: the tool must ensure high translation accuracy to reduce the need for manual corrections. Second, generating safe Rust programs is challenging. The tool must generate idiomatic, safe Rust code, ensuring safety and reliability in the resulting Rust code.

\begin{figure}[t]
\includegraphics[width=0.75\columnwidth]{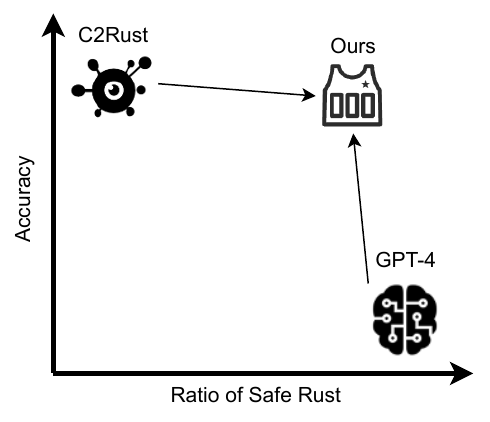}
\caption{The positioning of our approach relative to existing work. The x-axis represents the ratio of safe Rust code generated. The y-axis denotes the accuracy, or success rate, of transplanting automatically or semi-automatically translated code. C2Rust serves as a state-of-the-art rule-based translation tool, while GPT-4 represents a state-of-the-art learning-based translation approach.}
\label{fig:positioning}
\centering
\Description{positioning}
\end{figure}

Two primary approaches to C-to-Rust translation have been explored: rule-based and, more recently, LLM-based approaches (Figure~\ref{fig:positioning}). First, a \textbf{rule-based} approach includes tools such as C2Rust~\cite{c2rust}. These tools use syntactic and semantic analysis to map equivalent constructs between C and unsafe Rust. While they prioritize maintaining fidelity to the original code, they mainly produce unsafe Rust. Second, an  \textbf{LLM-based} approach leverages deep learning models, particularly large language models (LLMs) such as CodeX~\cite{DBLP:journals/corr/abs-2107-03374}, GPT-4~\cite{DBLP:journals/corr/abs-2303-08774}, and Llama 3~\cite{DBLP:journals/corr/abs-2407-21783}. These models are highly effective in generating idiomatic and safe Rust code. However, their ability to understand long program structures and maintain semantic consistency across large files and modules is often limited. For instance, applying GPT-4o to translate the ramfs module of the Linux kernel results in 75\% idiomatic and safe Rust lines, but the overall accuracy of the translation is low, requiring a huge number of human modifications. This discrepancy is primarily due to a ``laziness'' problem, where the model tends to omit large sections of the code, replacing them with brief comments instead of providing full translations (Figure~\ref{fig:laziness}).


To address the `laziness'' problem observed in models such as GPT-4o when processing long contexts, in this paper, we present \system, an LLM-based C-to-Rust translation tool. Our analysis demonstrates that translating shorter, individual functions within large modules significantly mitigates this issue (Figure~\ref{fig:laziness_prob}). Guided by these insights, \system\ decomposes extensive modules into smaller, manageable functions, ensuring that function lengths align with the thresholds shown in Figures~\ref{fig:lines} and ~\ref{fig:laziness_prob}. It then translates each function independently and reassembles the translated functions into a cohesive and fully functional Rust module.


To implement \system, we address three key challenges. First, directly translating individual functions using an LLM often results in low accuracy due to insufficient context. We need an effective strategy to split a module into individual functions while preserving essential contextual information, such as externally referenced variables and other function dependencies. Second, LLMs occasionally fail to translate complex functions due to their inherent limitations. Therefore, we need backup solutions. Third, translation can modify function interfaces and variable definitions, causing misalignment. We need a robust approach to reassemble individual functions into a cohesive Rust module, resolving interface and dependency inconsistencies.

\system\ addresses the preceding three challenges. First, it constructs a function call graph and employs program analysis to resolve external variables, ensuring each function's context includes all necessary dependencies. Second, \system\ incorporates two backup mechanisms for functions that fail to translate. It first employs a program repair component to correct errors in unsuccessful translations and then relies on human intervention for any remaining issues. Third, \system\ translates functions following the topology of the call graph, finalizing function interfaces translated earlier in the process. In addition, \system\ uses rule-based approaches to finalize other variable definitions.

We implement \system\ using program analysis tools, including tree-sitter~\cite{tree-sitter}, to perform function splitting and fusion, GPT-4o for translating individual functions, and rule-based tools such as C2Rust and BindGen to enhance the process. We evaluate \system\ by applying it to translate three representative Linux kernel modules: math, sort, and ramfs. On average, \system\ reduces human intervention to less than 15\% of the code lines.

In summary, this paper makes the following main contributions:

\begin{itemize}
    \item We identify and analyze the ``laziness'' problem in LLM-based approaches for translating large C modules with long contexts into Rust,  hindering their effectiveness in handling large-scale system translations.
    \item We propose \system, a C-to-Rust translation tool that directly addresses the ``laziness'' problem, aiming to accelerate the translation of complex systems while minimizing human intervention. 
    \item We evaluate \system\ on the Linux kernel, demonstrating its effectiveness by reducing the manual effort required to less than 15\% of the code lines.

\end{itemize}
\section{Background and Motivations}
\label{sec:background}

\subsection{The urgent need of code translation tools}

In recent years, there has been a growing trend toward rewriting applications in the Rust programming language~\cite{DBLP:conf/usenix/LiGYWX24, DBLP:conf/esem/PanterE24}. Rust’s memory safety guarantees and modern language features offer compelling advantages for system-level software, which traditionally relies on C. C, while performant and widely adopted, lacks inherent protections against common vulnerabilities such as buffer overflows and null pointer dereferences. Rust addresses these issues with a strict ownership model and compile-time checks, enabling safer and more robust software without compromising on performance. 

Despite these benefits, transitioning complex, large-scale codebases such as the Linux kernel, database systems, and compilers to Rust poses significant challenges. For instance, the Linux kernel represents one of the most intricate software projects in modern computing. Its design spans millions of lines of code across five major subsystems, interconnected through hundreds of interfaces. Over three decades of development, the kernel has grown to exceed 32 million lines of code as of 2021, contributed by more than 10,000 developers from over 1,200 organizations worldwide~\cite{DBLP:conf/esem/PanterE24}. This immense scale underscores the challenges of maintaining and translating such a sophisticated system.

One solution is to use a code translation tool that automates or semi-automates the process of converting existing C code into Rust. Such tools aim to bridge the gap between legacy codebases and modern Rust features, offering a more efficient path toward system-wide refactoring. Though the use of a translation tool cannot entirely remove the need for human oversight, it does alleviate much of the repetitive work, thereby significantly reducing the effort required.

\subsection{Limitations of existing code translation approaches}


There are two main approaches to achieving effective code translation: rule-based and LLM-based.

\paragraph{Rule-based approaches.} Rule-based approaches include tools such as transpilers~\cite{c2rust, bindgen, CxGo}, or source-to-source compilers, that employ program analysis techniques to convert code between programming languages. \textbf{Advantage}: Rule-based approaches excel in preserving the structure and semantics of the original code, producing reliable and consistent output that is essential for system-level translations. \textbf{Limitations}: These approaches encounter significant challenges when dealing with complex language features or idioms that lack direct counterparts in the target language, such as translating C pointer arithmetic into safe Rust references. For example, C2Rust~\cite{c2rust}, the state-of-the-art rule-based C-to-Rust translation tool, generates only unsafe Rust. While this approach achieves high translation accuracy, it fails to leverage Rust's safety guarantees, undermining one of the key motivations for migrating to Rust. BindGen~\cite{bindgen}, however, can generate safe Rust but only for macro or type definitions.

\begin{figure}[t]
\includegraphics[width=\linewidth]{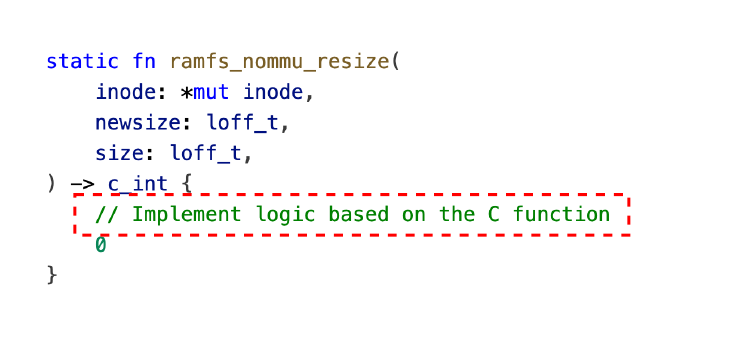}
\caption{Translated Rust code snippets of the ramfs module, where the LLM omits essential details (``laziness'' problem).}
\centering
\label{fig:laziness}
\Description{laziness}
\end{figure}

\paragraph{LLM-based approaches.} LLM-based approaches represent the current state-of-the-art in deep learning for code translation, leverage Large Language Models (LLMs) such as CodeX~\cite{DBLP:journals/corr/abs-2107-03374}, GPT-4~\cite{DBLP:journals/corr/abs-2303-08774}, and Llama 3~\cite{DBLP:journals/corr/abs-2407-21783}. \textbf{Advantage}: LLM-based approaches can learn to translate diverse language features and idioms, generating more idiomatic code in the target language, including safe Rust. \textbf{Limitations}: Due to their data-driven nature, LLMs may occasionally produce syntactically or semantically incorrect code, particularly for long or complex translation tasks. For example, GPT-4o, the state-of-the-art LLM exhibits a ``laziness'' problem that hinders its reliability. In such cases, GPT-4o may omit significant portions of the original code to pass Rust's stringent compiler checks (Figure~\ref{fig:laziness}).





\section{Challenges and Design}

\begin{figure}[t]
    \centering
    \includegraphics[width=0.9\linewidth]{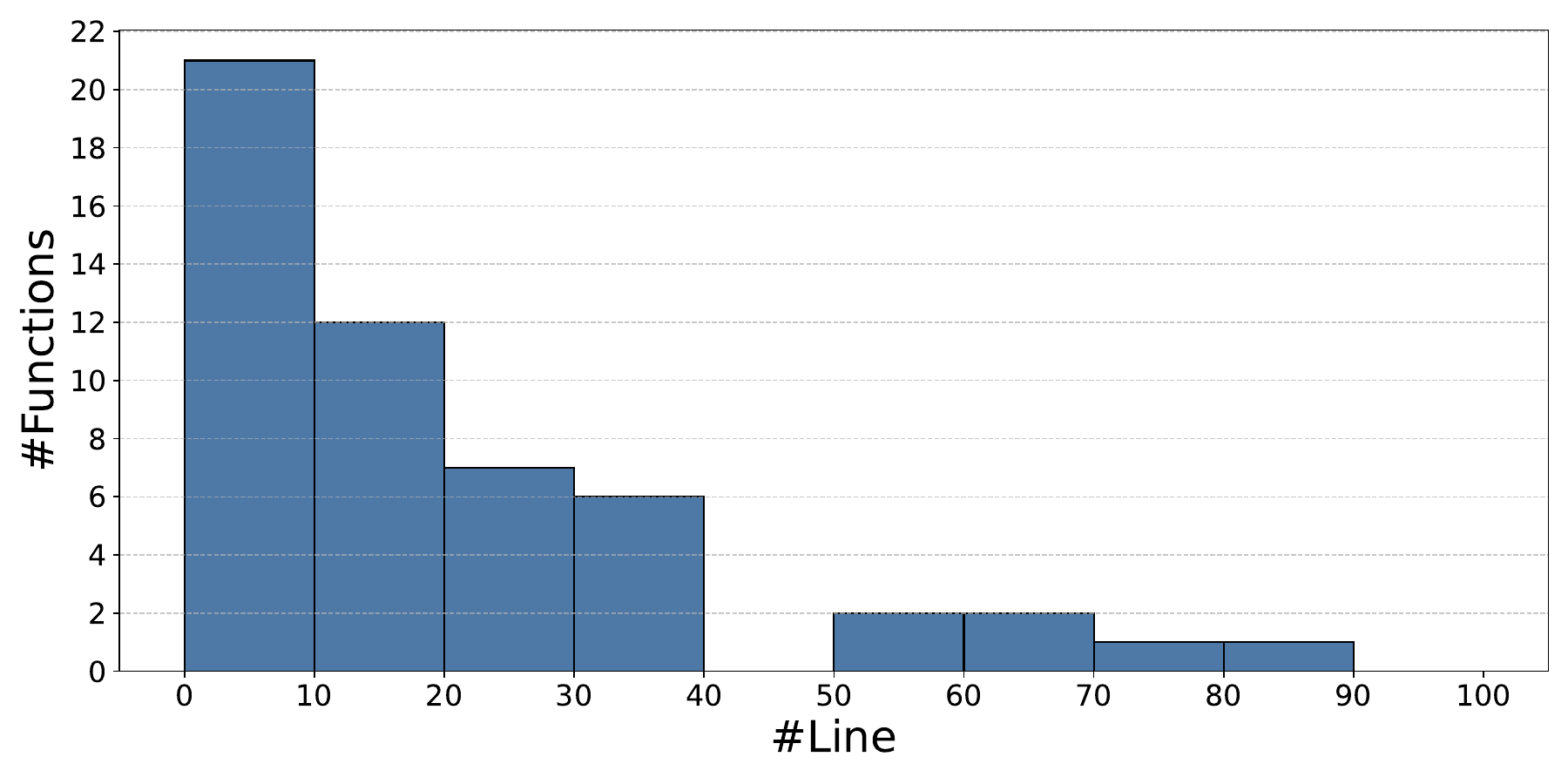}
    \caption{The distribution of numbers of code lines (\# Line) per function across three modules (math, sort, ramfs) in the Linux kernel. The total number of lines of code for all functions is 496 in the math module, 173 in the sort module, and 379 in the ramfs module.}
    \label{fig:lines}
    \Description{lines}
\end{figure}

\begin{figure}[t]
    \centering
    \includegraphics[width=\linewidth]{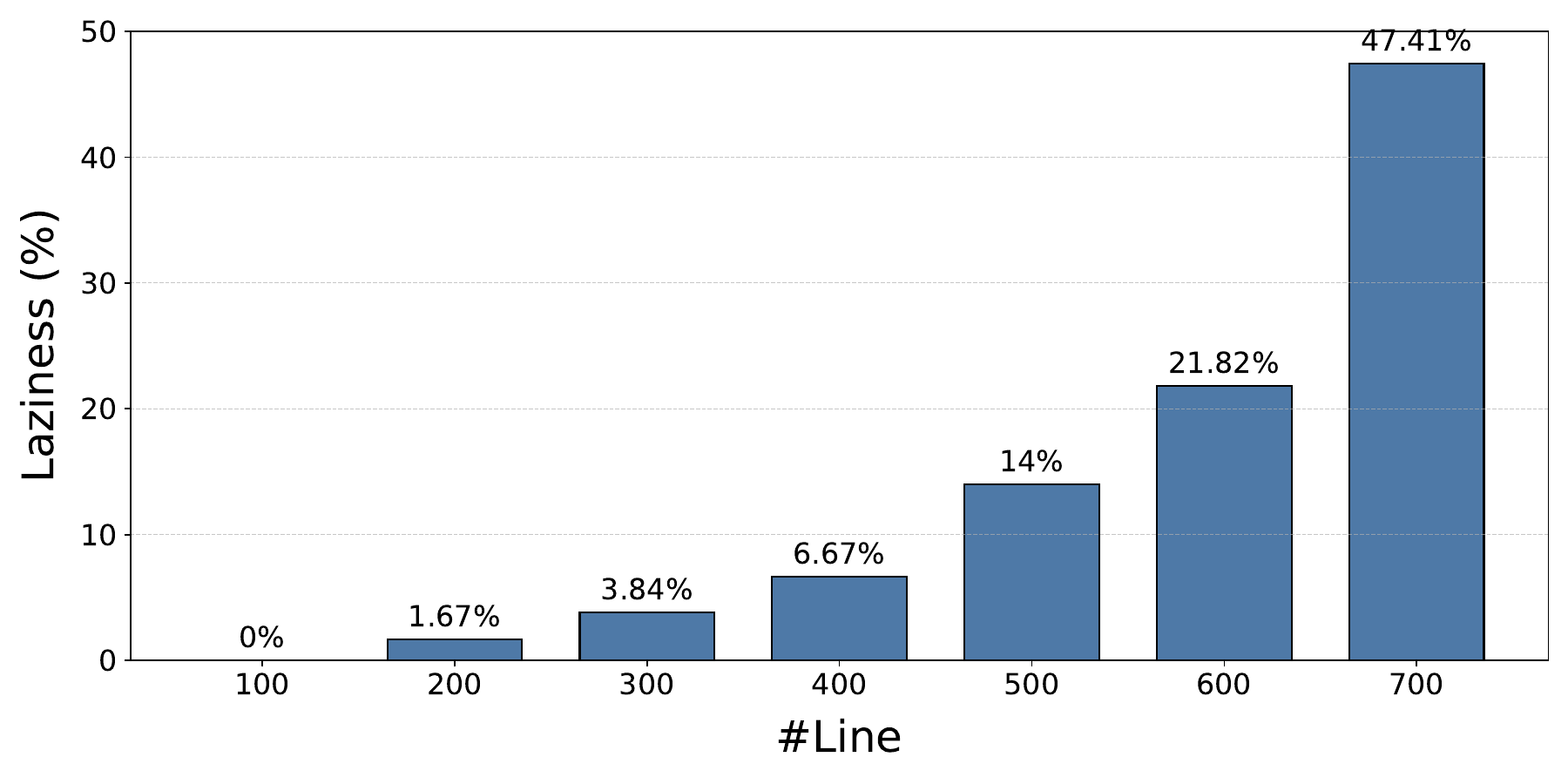}
    \caption{The Probability of Laziness Across Different \# Line. As the number of code lines (\# Line) increases, the probability of laziness exhibits a rising trend.}
    \label{fig:laziness_prob}
    \Description{laziness}
\end{figure}

The ``laziness'' problem refers to the tendency of the language model to omit significant portions of code, sometimes replacing them with brief comments rather than providing full translations, as illustrated in Figure~\ref{fig:laziness}. Specifically, we define ``laziness'' by checking whether the model completely replaces or skips any substantial code blocks during the translation. As shown in Figure~\ref{fig:laziness_prob}, we measure the percentage of translations that exhibit this omission behavior. For example, when translating smaller segments of code below 200 lines, GPT-4o exhibits ``laziness'' in only 1.67\% of cases, whereas for segments of about 700 lines, the omission rate jumps significantly to 47.4\%. These findings reveal that the likelihood of omissions correlates strongly with the length of the code segment.

We have observed this phenomenon when translating large Linux kernel modules, where the model struggles to process, understand, and fully translate lengthy code. Through further analysis, we observe that the size of a single function, in terms of lines of code, is typically moderate, as shown in Figure~\ref{fig:lines}. Therefore, we adopt a strategy of dividing the translation task into smaller, more manageable units --- individual functions --- and translating each function independently. Our investigation reveals that this function-level translation approach significantly mitigates the ``laziness’’ problem, resulting in more accurate, complete, and reliable translations, as demonstrated in Figure~\ref{fig:laziness_prob}.

\begin{figure*}[t]
    \centering
    \includegraphics[width=1\linewidth]{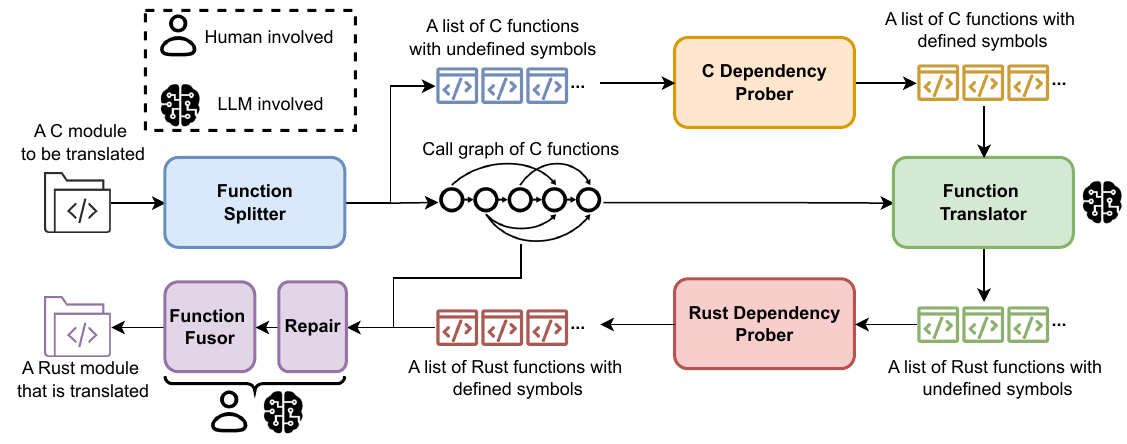}
    \caption{System overview.}
    \label{fig:sys}
    \Description[fig:sys]{sys}
\end{figure*}

Based on these findings, we introduce \system, an LLM-based C-to-Rust translation tool that addresses the ``laziness'' problem. As shown in Figure~\ref{fig:sys}, \system\ decomposes large code modules into smaller, more manageable units --- individual functions, translates each function independently, and then reassembles them to form a complete Rust module.

\subsection{Challenges}
\label{sec:challenges}

To implement \system, we must address three key challenges, as outlined in the following three sections.

\begin{figure}[t]
    \centering
    \includegraphics[width=\linewidth]{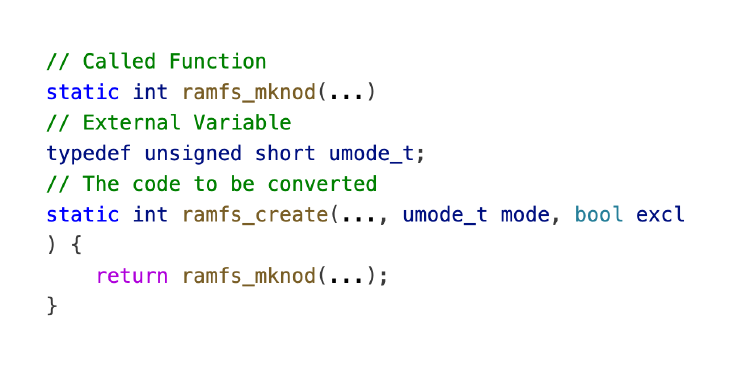}
    \caption{One splitted C function from the ramfs module, along with its contextual dependencies including external variables and functions it invokes.}
    \label{fig:function_split}
    \Description[fig:function\_split]{function_split}
\end{figure}

\paragraph{Challenge 1: Insufficient context information.} 

Directly translating individual functions using an LLM often results in low accuracy due to insufficient
context. As illustrated in Figure~\ref{fig:function_split}, while we aim to direct the LLM to translate only the core function --- ``the code to be converted'' part of the code --- the absence of relevant contextual information makes it difficult to produce an accurate translation. Although we can easily extract individual functions from a module, their contextual information is often dispersed across the entire codebase, necessitating an approach to identify and resolve these externally referenced symbols for a given core function.

\paragraph{Challenge 2: LLMs' inability to translate complex functions.} 

After decomposing a module into individual core functions and complementing them with contextual information (as illustrated in Figure~\ref{fig:function_split}), the next challenge is to translate these functions in the presence of LLMs' inability to translate complex ones. Despite extensive efforts, the LLM may occasionally fail to translate a core function, or the translated function might still fail to compile. This issue often arises from the inherent limitations of the LLM for understanding idiomatic requirements specific to Rust. For instance, certain Rust structures must implement traits like Copy, which lack direct counterparts in C. We need a backup mechanism to ensure successful translation and compilation when the LLM reaches its limits.




\paragraph{Challenge 3: Unalignment of function interfaces and variable definitions.}


The final step involves reassembling the individually translated Rust functions into a complete and functional Rust module, requiring alignment of contexts across core functions, including function interfaces and variable definitions. Since these contexts are shared across multiple core functions, translating each function independently may lead to inconsistencies. To address this challenge, we require an approach that separates the translation of contexts from the core functions and later integrates them to ensure consistency and correctness.



\subsection{System Design}
\label{sec:design}

To address the challenges outlined in the previous section, we propose the design of our system, as illustrated in Figure~\ref{fig:sys}. \system\ automates the translation of code modules written in C (a collection of source and header files) into their equivalent Rust implementations (a collection of source files) through the following five sequential steps. First, the \textit{Function Splitter} processes the C module to extract all functions from the provided source files and construct the module’s call graph, capturing the relationships and dependencies between functions. Second, the \textit{C Dependency Prober} resolves undefined symbols in the extracted functions by globally searching the original codebase (written in C) and adding appropriate declarations to the context of corresponding functions. Third, the \textit{Function Translator} leverages a Large Language Model (LLM) to translate individual C functions into Rust equivalents. Fourth, the \textit{Rust Dependency Prober} applies a compiler-driven approach to resolve undefined symbols in each translated Rust function. Fifth, the \textit{Function Fusor} integrates the translated Rust functions into a cohesive Rust module, following the topological order defined by the call graph.

In the following sections, we detail how \system\ addresses the challenges in Section~\ref{sec:challenges}.

\paragraph{Keep important contextual information when splitting core functions.}

\system\ addresses the challenge of resolving contextual information using two key components: the \textit{Function Splitter} and the \textit{C Dependency Prober}. First, the \textit{Function Splitter} extracts individual functions from the source files of the C module, saving each function in a separate file. These extracted functions may contain undefined symbols, such as variables or function invokes (``The code to be converted'' part in Figure~\ref{fig:function_split}). Simultaneously, the \textit{Function Splitter} constructs a call graph to represent the relationships and dependencies among functions in the module. Second, the \textit{C Dependency Prober} leverages the call graph to identify external functions and conducts a global search within the codebase to locate variable definitions. Then, it appends declarations of each symbol to the corresponding functions excluding the full implementations (Figure~\ref{fig:function_split}). The exclusion of full implementations helps reduce the size of the prompt passed to the LLM, minimizing the risk of triggering the ``laziness'' problem.


\begin{figure}[t]
    \centering
    \includegraphics[width=1\linewidth]{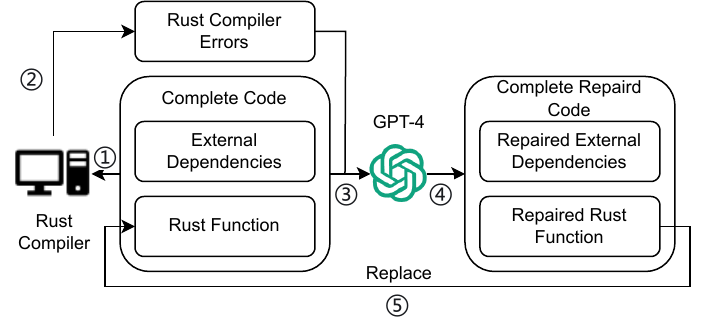}
    \caption{Program repair workflow.}
    \label{fig:repair}
    \Description[figrepair]{repair}
\end{figure}

\paragraph{Solutions for LLMs' inability to translate complex functions.}

\system\ integrates a dedicated \textit{Repair} component into its workflow, positioned after the \textit{Rust Dependency Prober} and before the \textit{Function Fusor} (Figure~\ref{fig:repair}). This component provides an additional opportunity for the LLM to rectify errors in the translated code. We hypothesize that repair tasks are simpler than generation tasks, thereby increasing the likelihood of success. Empirical results demonstrate that within three iterations of repair, the LLM-generated code successfully compiles in most cases (Table~\ref{tab-new-repair}).

The repair process proceeds as follows. First, we compile the code and capture errors reported by the Rust compiler. Second, we combine the compiler error messages with the problematic code and provide this input to the LLM. Third, we instruct the LLM to repair the provided code. Fourth, we accept only the repaired core function while keeping the external dependencies untouched, which were primarily generated by rule-based tools like BindGen. Finally, we replace the original function with the repaired one.

If the LLM fails to repair the function within three attempts, we abandon the LLM-based approach and resort to rule-based solutions, such as C2Rust~\cite{c2rust}. This step balances translation accuracy, and the proportion of safe Rust code while ensuring successful translation (Figure~\ref{tab:e2e}).

\begin{figure}[t]
    \centering
    \includegraphics[width=\linewidth]{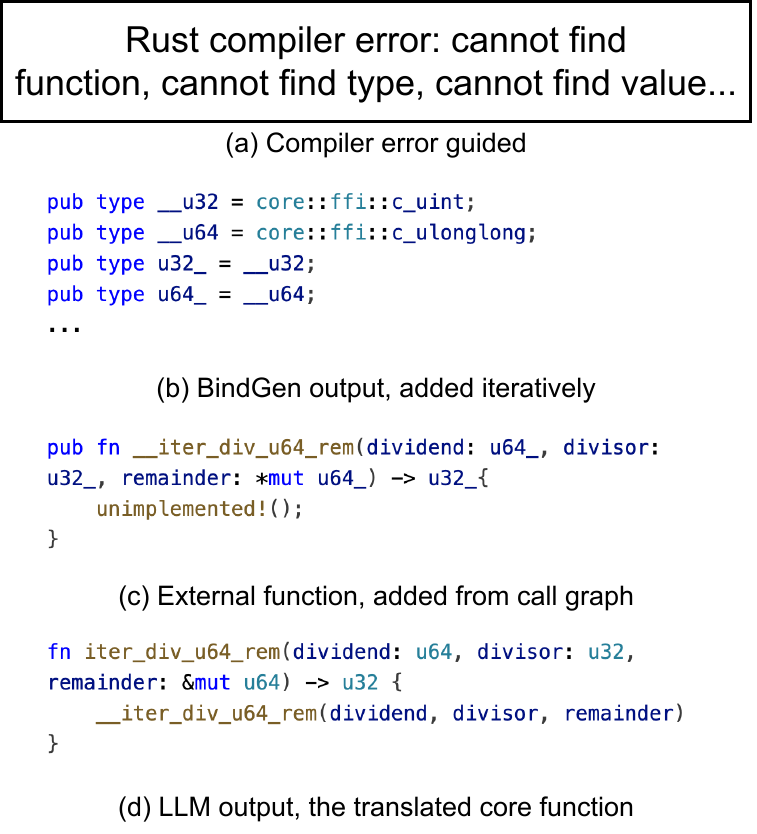}
    \caption{The translation process of one C function from the ramfs module. BindGen translates the definitions of its external variables, while LLM translates the C function and its invoked external functions into Rust. The Rust Dependency Prober iteratively integrates these translations, resolving any undefined symbols and ensuring correctness through a compiler-driven process.}
    \label{fig:function_translator}
    \Description[fig:function\_translator]{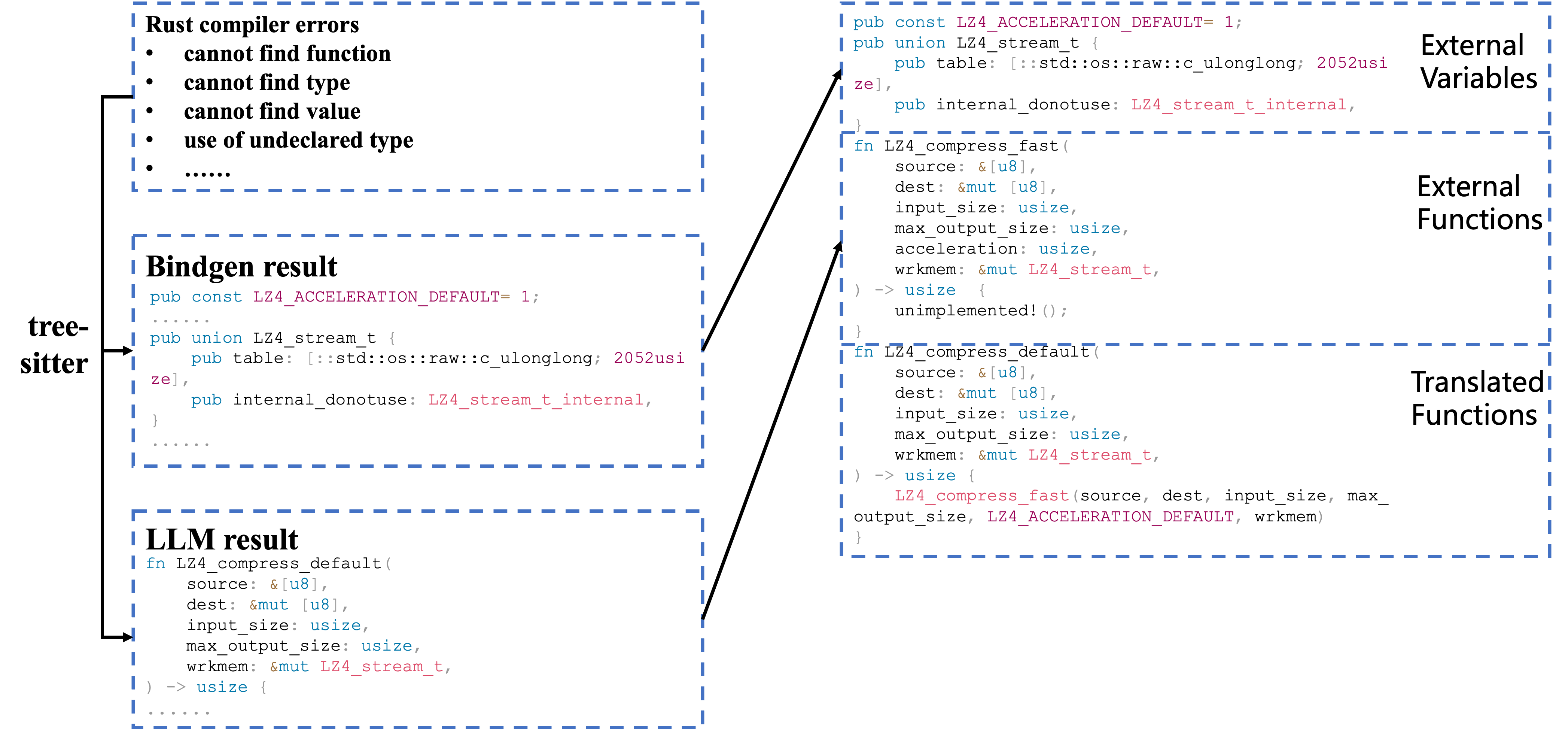}
\end{figure}

\paragraph{Align the contexts of core functions.}

\system\ effectively separates and resolves the translation of contexts of core functions using two components: the \textit{Function translator} and the \textit{Rust Dependency Prober}. First, the \textit{Function Translator} employs a Large Language Model (LLM) to translate individual core functions into Rust, augmented by their contexts. It follows the topological order defined by the function call graph. However, it retains only the translated core function while discarding translated external symbols, including function declarations and variable definitions. Concurrently, the \textit{Function Translator} uses BindGen~\cite{bindgen}, a rule-based tool, to generate consistent definitions for external variables such as structures and macros. Second, the \textit{Rust Dependency Prober} iteratively resolves undefined external symbols in the translated Rust functions (Figure~\ref{fig:function_translator}). For external function declarations, it utilizes the module’s call graph to locate and incorporate corresponding translated external functions. For external variables, it searches the BindGen output, appending the necessary definitions of structures, macros, and others. This component adopts a compiler-driven approach, involving repeated cycles of code compilation, identification of unresolved symbols, and incorporation of missing definitions. The process continues iteratively until all external dependencies are resolved, yielding a complete and compilable Rust module (Figure~\ref{fig:function_translator}).

Thanks to the preceding two components, at the final fusing stage, \system\ employs a simple component, the \textit{Function Fusor}, to assemble functions into a cohesive and compilable Rust module based on the topological order of the function call graph. The \textit{Function Fusor} begins with leaf functions (those that do not invoke other functions), removes duplicate external dependencies between the leaf node and its parent, and integrates the leaf node’s code into its parent function. This process iterates systematically through the function call graph in topological order, fusing functions layer by layer until all nodes are combined into a single, cohesive Rust module. Any residual errors encountered during this stage are resolved manually, leveraging program repair tools whenever feasible. By consolidating all required human intervention into this final stage, \system\ ensures a streamlined and efficient workflow for generating a fully functional Rust module.

\section{Implementation}
\label{sec:impl}

\system\ comprises three types of components: those utilizing a Large Language Model (LLM), those involving human interaction, and those based on rule-driven tools. For LLM-based components, we interact with the GPT-4o model via the OpenAI API. The human-interaction components feature a text-based user interface, enabling users to assist in correcting code. Rule-based tools, such as Tree-sitter~\cite{tree-sitter} for function splitting, slicing, and fusion, and Bindgen~\cite{bindgen} for generating definitions of external variables, such as structures and macros, are employed for other tasks. This section provides additional implementation details not covered in Section~\ref{sec:design}.

\paragraph{Prompt templates for \textit{Function Translator}.}

\begin{figure}[t]
    \centering
    \includegraphics[width=\linewidth]{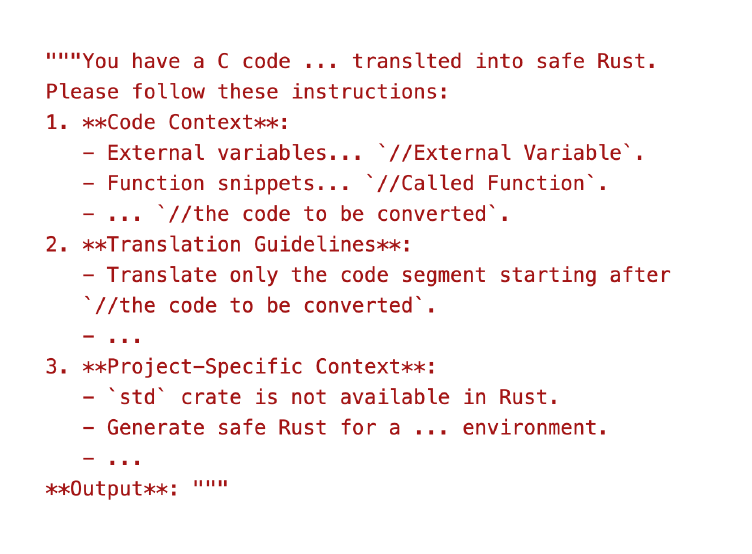}
    \caption{The system prompt for \textit{Function Translator}.}
    \label{fig-translator-prompt}
    \Description[fig-translator-prompt]{fig-translator-prompt}
\end{figure}

For the \textit{Function Translator}, we use the prompt illustrated in Figure~\ref{fig-translator-prompt}, comprising three key components. First, we define the code context, specifying that the code includes ``external variables,'' ``called functions,'' and the core function targeted for translation. Second, we outline translation-specific guidelines, instructing the LLM to focus exclusively on translating the core function. Third, we detail project-specific requirements, such as excluding the use of the ``std'' crate, which is unavailable in the environment.

\paragraph{Ensuring syntactic correctness in \textit{Function Translator}.}

LLM-generated functions often contain syntax errors, necessitating a robust mechanism to ensure syntactic correctness in the translated output. \system\ addresses this issue through iterative refinement within the \textit{Function Translator}. It employs a syntax parser, such as Tree-sitter~\cite{tree-sitter}, to verify correctness after each translation attempt. If the generated Rust function contains syntax errors, the system prompts the LLM to retry the translation iteratively until it produces a valid output. Empirical evidence demonstrates that the LLM typically generates a syntactically correct function within two to three iterations. This step is particularly susceptible to the ``laziness'' problem, where the LLM, when tasked with translating long functions, tends to produce incomplete results that only superficially pass validation checks rather than addressing all required details comprehensively. However, our function-level translation design mitigates this issue effectively, ensuring that the LLM generates fully correct outputs.

\paragraph{Prompt templates for \textit{Repair}.}
\begin{figure}[t]
    \centering
    \includegraphics[width=\linewidth]{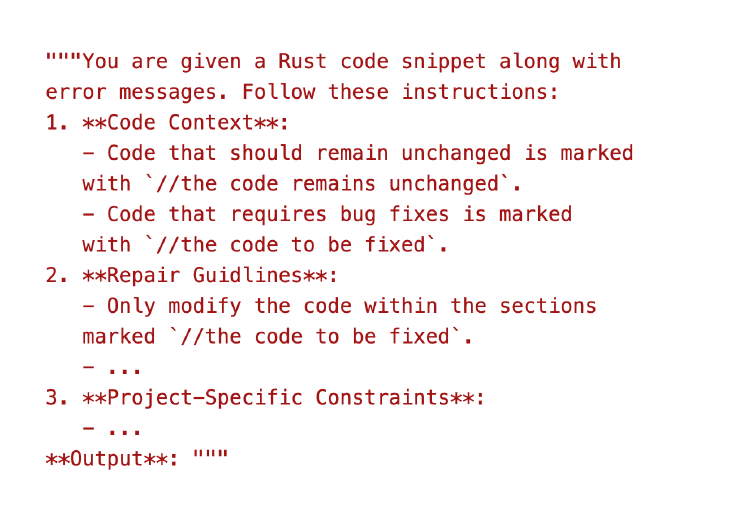}
    \caption{The system prompt for \textit{Repair}.}
    \label{fig-repair-prompt}
    \Description[fig-repair-prompt]{fig-repair-prompt}
\end{figure}

For the \textit{Repair}, we use the prompt shown in Figure~\ref{fig-repair-prompt}, which also consists of three key components, similar to the prompt for the Function Translator. The primary difference lies in the code context and the inclusion of repair-specific guidelines.

\paragraph{Number of iterations for program repair.}

The \textit{Repair} process undergoes a fixed number of iterations (e.g., three in our implementation). If the function still fails to compile after these iterations, we manually address the errors. For particularly difficult errors, we substitute the Rust function with unsafe Rust code generated by C2Rust~\cite{c2rust}, ensuring that the module remains functional.

\paragraph{\textit{Function Fusor} details.}

During the final step of function fusion, we merge functions according to the topological order defined by the call graph. Each time we fuse a leaf node into its parent function, we compile the partially merged code to ensure compatibility. This process continues iteratively until the entire module is fused. Once the module passes compilation, we integrate it into the original C-based codes. We then compile the kernel and verify that it runs successfully. Only after these steps do we consider the translation of the module complete.

\section{Evaluation}
\label{sec:eval}

In this section, we explore three research questions (RQs) to evaluate the effectiveness of \system\ in comparison to traditional rule-based and LLM-based approaches:

\begin{itemize} 
\item RQ1: What is the amount of human effort required to achieve a fully functional translation (100\% functional accuracy), and what proportion of the resulting code adheres to Rust's safety principles when using \system?
\item RQ2: What types of errors require manual intervention, and how many such errors are encountered during the translation process?
\item RQ3: What are the advantages of function-level translation compared to whole-module translation?
\item RQ4: How does the \textit{Repair} component contribute to the translation process?
\end{itemize}

\subsection{Setup}

\paragraph{Dataset.}

\begin{table}[ht]
\begin{tabular}{ccccc}
\toprule
{\color[HTML]{333333} Dataset} &
  {\color[HTML]{333333} \# Funs} &
  {\color[HTML]{333333} \# Lines} &
  {\color[HTML]{333333} \begin{tabular}[c]{@{}c@{}}\# Extern\\ vars\end{tabular}} &
  {\color[HTML]{333333} \begin{tabular}[c]{@{}c@{}}\# Extern\\ funs\end{tabular}} \\ \midrule
{\color[HTML]{333333} math}   & {\color[HTML]{333333} 18} & {\color[HTML]{333333} 809} & {\color[HTML]{333333} 12} & {\color[HTML]{333333} 19} \\ \hline
{\color[HTML]{333333} sort}  & {\color[HTML]{333333} 12} & {\color[HTML]{333333} 291}  & {\color[HTML]{333333} 12} & {\color[HTML]{333333} 8}  \\ \hline
{\color[HTML]{333333} ramfs} & {\color[HTML]{333333} 21} & {\color[HTML]{333333} 637}  & {\color[HTML]{333333} 31} & {\color[HTML]{333333} 53} \\ \bottomrule
\end{tabular}
\caption{Statistics of datasets. Each dataset is an example module from the Linux kernel with a varying number of functions, lines of code, external variables, and external functions.}
\label{tab:datasets}
\end{table}

We select the following three Linux kernel modules for translation using \system\ with human assistance, based on two key considerations. First, the translation of the selected modules from C to Rust is urgent due to a tight project schedule. We need to finish their translation within the available time. These three modules are prioritized as they are part of the current agenda. Second, the translation process relies on manual effort, which limits the scalability of our testing approach. As a result, we focus on translating and testing only three modules to ensure the feasibility of completing the task within the given constraints. The specific data of datasets is shown in Table~\ref{tab:datasets}.

\begin{itemize}
    \item \textbf{math}: The math module provides a collection of efficient, integer-based mathematical functions designed to support various kernel operations with minimal computational overhead.
    \item \textbf{sort}: The sort module in the Linux kernel offers a generic and reusable implementation of a sorting algorithm, enabling various kernel components to efficiently sort data. 
    \item \textbf{ramfs}: The ramfs (RAM File System) module in the Linux kernel is a lightweight, in-memory file system designed to store files and directories directly in RAM. Due to the larger contextual information associated with each core function, the translation process for this module is more complex compared to others.
\end{itemize}

\paragraph{Baseline.}

We choose \textbf{GPT-4o} as our baseline. GPT-4o is a multilingual, multimodal generative pre-trained transformer developed by OpenAI and released in May 2024. GPT-4o can process and generate text, images, and code, which makes it a good candidate for automatic code translation (from C to Rust specifically).

\paragraph{Environment Setup.}

We implement \system\ using Python 3.8 and Tree-sitter 0.21.1. Tree-sitter~\cite{tree-sitter} is a parsing library designed to build fast and efficient incremental parsers for programming languages. Additionally, we use Bindgen 0.56.0 and C2Rust 0.18.0 in our implementation. We conduct our experiments on a machine equipped with a 3.8 GHz AMD Ryzen 7 8845H CPU, which has 8 physical cores and 16 hyper-threads. The system is configured with 32 GB of RAM and runs Ubuntu 20.04.

\paragraph{Metrics.}


We select the following metrics.

\begin{itemize}
    \item \textbf{Manually Modified Lines (MML).} We use \#MML to denote the absolute number of manually modified lines, and \%MML to represent the proportion of these lines relative to the total number of lines in the translated code. This metric quantifies the extent of manual intervention required to make the project fully run and correctly function with the translated modules. In other words, we make our best effort to ensure that each translated module not only compiles but also runs as expected before counting any modifications toward MML. Consequently, MML reflects the essential adjustments that remain in the final version of the code once it is verified to compile and pass functional tests. Additionally, we define MML-LLM to measure the manual modifications required specifically for LLM-translated code to achieve successful compilation and operation.
    \item \textbf{Safe Code (SC).} We use \#SC to denote the number of lines of safe Rust code within the generated Rust code, specifically the lines not enclosed by the unsafe keyword. We use \%SC to represent the proportion of these lines relative to the total number of lines in the translated code. This metric reflects the extent to which the translated code adheres to Rust's memory safety guarantees.
   
   \item \textbf{Compilation Success Rate (CSR).} CSR measures the compilability of the generated code by calculating the proportion of functions that compile successfully. It is defined as the ratio of the number of functions that pass compilation to the total number of generated code samples. This metric indicates the effectiveness of the translation process in producing syntactically correct and compilable Rust code.
   \item \textbf{Laziness.} Laziness refers to the proportion of functions in the translation results that exhibit an omission of complete translation, indicating the ``laziness'' problem of LLMs.
    \item \textbf{CodeBLEU (C-BLEU).}
    The CodeBLEU~\cite{DBLP:journals/corr/abs-2009-10297} score, adapted from BLEU and designed specifically for programming language processing, evaluates the accuracy of code generation models.




\end{itemize}


\subsection{RQ1. The amount of human labor and the proportion of safe Rust.} 

\begin{table*}[ht]
\begin{tabular}{c|cc|ccc|ccc}
\toprule
{\color[HTML]{333333} Dataset} & 
{\color[HTML]{333333} \# Line} & 
{\color[HTML]{333333} \# Line-LLM} & 
{\color[HTML]{333333} \# MML} & 
{\color[HTML]{333333} \# SC} & 
{\color[HTML]{333333} \# SC-LLM} & 
{\color[HTML]{333333} \% MML} & 
{\color[HTML]{333333} \% SC} & 
{\color[HTML]{333333} \% SC-LLM} \\ 
\midrule
{\color[HTML]{333333} math} & 
{\color[HTML]{333333} 570} & 
{\color[HTML]{333333} 535} & 
{\color[HTML]{333333} 74} & 
{\color[HTML]{333333} 564} & 
{\color[HTML]{333333} 529} & 
{\color[HTML]{333333} 12.98\%} & 
98.94\% & 
98.88\% \\ 
\midrule
{\color[HTML]{333333} sort} & 
{\color[HTML]{333333} 321} & 
{\color[HTML]{333333} 293}  & 
{\color[HTML]{333333} 46} & 
{\color[HTML]{333333} 134} & 
{\color[HTML]{333333} 106} & 
{\color[HTML]{333333} 14.33\%} & 
41.74\% & 
36.18\% \\ 
\midrule
{\color[HTML]{333333} ramfs} & 
{\color[HTML]{333333} 5803} & 
{\color[HTML]{333333} 421} & 
{\color[HTML]{333333} 277} & 
{\color[HTML]{333333} 4786} & 
{\color[HTML]{333333} 265} & 
{\color[HTML]{333333} 4.77\%} &
82.47\% & 
62.95\% \\ 
\bottomrule
\end{tabular}
\centering
\caption{End-to-end results of translating three datasets using \system. For each dataset, we report the number of final code lines (\# Line) after translation and the number of LLM-generated code lines (\# Line-LLM). The final code consists of LLM-generated code augmented with Bindgen-generated code, external functions, and human modifications and we report the number and proportion (\%) of Manually Modified Lines (MML) during this augmentation process. We also report the number and proportion of Safe Code (SC) in the final code, and in the LLM-generated code.}
\label{tab:e2e}
\end{table*}

Table~\ref{tab:e2e} summarizes the end-to-end results of using \system\ to translate three Linux kernel modules, followed by human modifications to make the translated module run successfully. Two key observations emerge from these results. First, despite the translated version of the three modules collectively containing hundreds to thousands of lines of code, \system\ reduces the human effort required for translation to less than 15\% of the final code lines. Second, unlike traditional rule-based tools such as C2Rust, \system\ generates safe Rust code, with up to 98.9\% of the translated code being safe, and 41.7\% in the worst case.

Although the numbers in the table highlight \system's capability to generate memory-safe Rust code with minimal human intervention, we also have the following two unusual observations. First, the sort module exhibits the lowest proportion of safe code. Its main functionality is implemented within only 12 functions, leading to a large ratio of unsafe code when even a few functions are unsafe. Second, ramfs has the smallest \% MML because its \# Line is disproportionately large. This inflated line count results from Bindgen generating complex contexts, such as headers and macros, due to ramfs's complexity and numerous dependencies.

\begin{figure}[t]
    \centering
    \includegraphics[width=\linewidth]{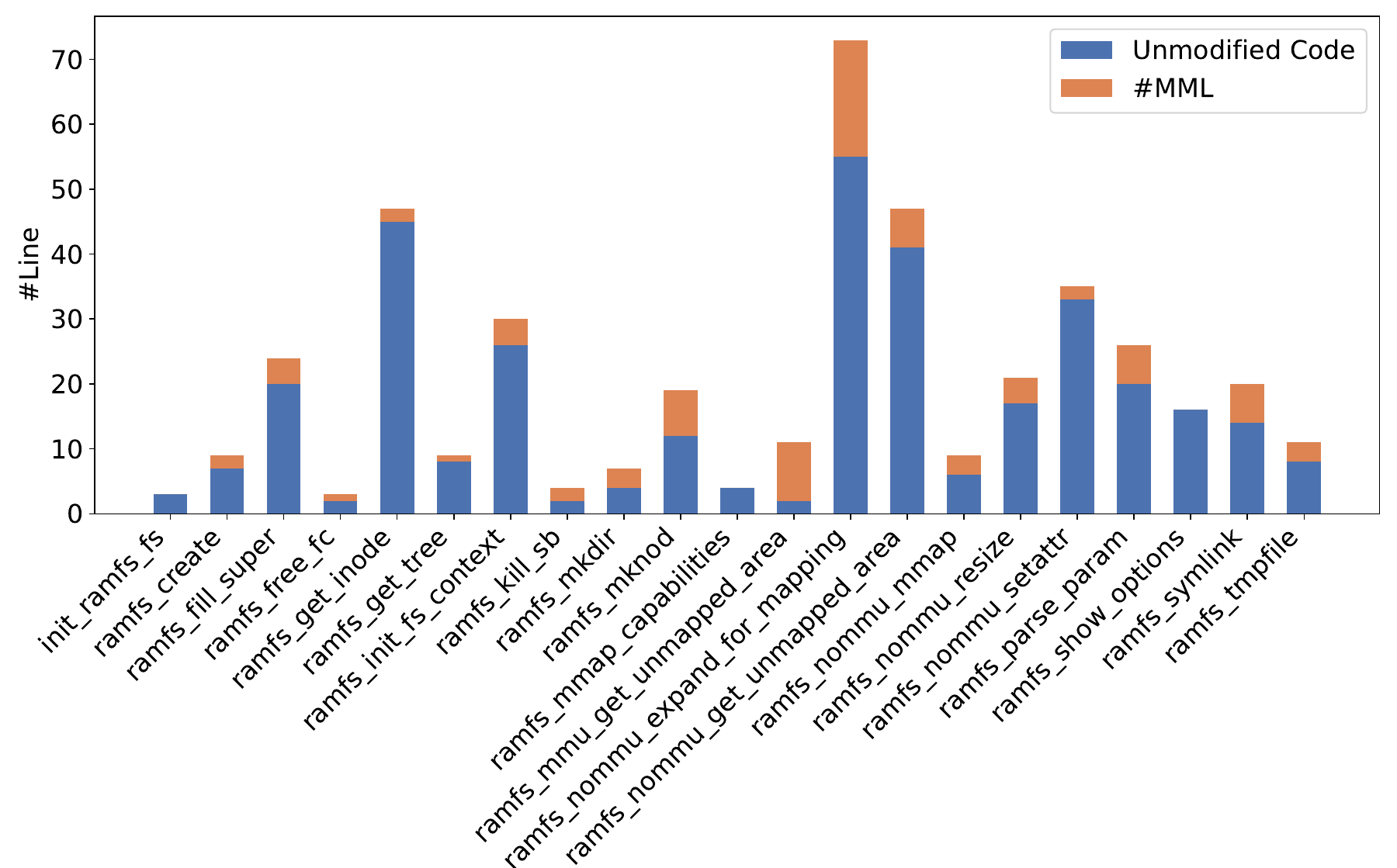}
    \caption{The breakdown of MML for each function from the ramfs module.}
    \label{fig-eval-function-mml}
    \Description[fig-eval-function-mml]{fig-eval-function-mml}
\end{figure}

Figure~\ref{fig-eval-function-mml} presents the breakdown of MML for each function in the ramfs module. We observe that MML is largely proportional to the function's line count. This correlation is expected, as longer functions are inherently more complex, making them more challenging for LLMs to translate accurately.

\subsection{RQ2. The types and frequency of errors that require manual intervention}

\begin{table}[ht]
\begin{tabular}{cccccc}
\toprule
{\color[HTML]{333333} \begin{tabular}[c]{@{}c@{}}\underline{Error}\\ Dataset\end{tabular}} &
  {\color[HTML]{333333} TE} &
  {\color[HTML]{333333} MI} &
  \begin{tabular}[c]{@{}c@{}}CS\end{tabular} &
  {\color[HTML]{333333} SE} &
  Others \\ \midrule
{\color[HTML]{333333} math} & {\color[HTML]{333333} 13} & {\color[HTML]{333333} 2} & 2 & 0 & 0  \\ \midrule
sort                        & 5                        & 24                        & 0 & 1 & 4 \\ \hline
ramfs                       & 140                      & 5                        & 6 & 0 & 4  \\ \bottomrule
\end{tabular}
\centering
\caption{The types and frequency of errors that require manual resolution when using \system to translate three datasets.}
\label{tbl-error}
\end{table}

\begin{figure}[t]
    \centering
    \includegraphics[width=\linewidth]{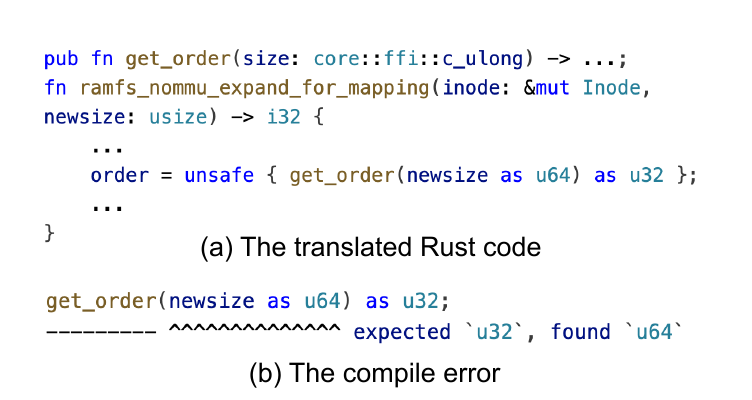}
    \caption{Type error example: type mismatch.}
    \label{fig-te}
    \Description{te}
\end{figure}

\begin{figure}[t]
    \centering
    \includegraphics[width=\linewidth]{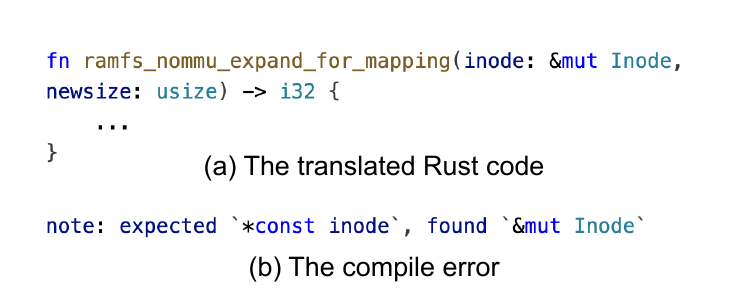}
    \caption{Changed symbol example: a symbol that does not exist.}
    \label{fig-cs}
    \Description{cs}
\end{figure}

\begin{figure}[t]
    \centering
    \includegraphics[width=\linewidth]{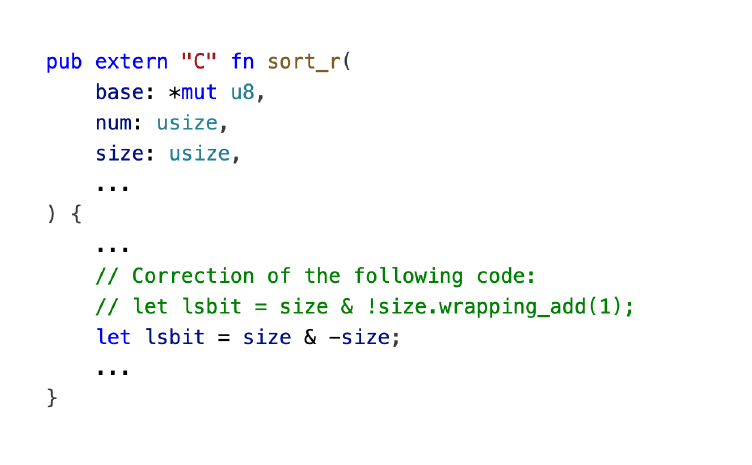}
    \caption{Semantic error example.}
    \label{fig-sem}
    \Description{sem}
\end{figure}

\begin{table*}[ht]
\begin{tabular}{cccccccccc}
\toprule
\multirow{2}{*}{Methods} &
  \multicolumn{3}{c}{\#MML} &
  \multicolumn{3}{c}{CodeBLEU} &
  \multicolumn{3}{c}{Laziness} \\ \cline{2-10} 
 & math & sort & ramfs & math & sort & ramfs & math & sort & ramfs \\ \midrule
GPT-4o & 266 & 127 & 5686 & 0.5303 & 0.4333 & 0.2679 & 32.22\% & 25\% & 76.67\% \\ \midrule
\system & 74 & 46 & 277 & 0.8707 & 0.5568 & 0.4666 & 5.56\% & 0\% & 0\% \\ \bottomrule
\end{tabular}
\caption{Performance comparison between GPT-4o and \system\ on translating three Linux kernel modules: math, sort, and ramfs. The evaluation includes three metrics: Manually Modified Lines (\#MML), CodeBLEU, and Laziness. MMLs for GPT-4o are estimated based on the difference between the ground truth Rust code lines and the lines generated by GPT-4o.}
\label{tab:function-level}
\end{table*}

\begin{table}[ht]
\begin{tabular}{cccccccc}
\toprule
RR & E0425 & E0308 & E0689 & E0574 & E0573 & CSR \\ \midrule
0  & 30    & 2    & 2    & 2    & 2    & 7/18  \\ \midrule
1  & 9    & 5    & 0    & 0    & 2    & 12/18  \\ \midrule
2  & 8    & 1    & 0    & 0    & 2    & 14/18  \\ \midrule
3  & 8    & 0    & 0    & 0    & 2    & 15/18  \\ \bottomrule
\end{tabular}
\centering
\caption{The impact of program repair following the Rust Dependency Prober in addressing LLMs' inability to translate complex functions. It presents the distribution of error types, categorized by error codes, encountered during the translation of 18 functions in the math module. Additionally, it shows the progression in the Compilation Success Rate (CSR) functions after zero, one, two, and three rounds of repair (RR).}
\label{tab-new-repair}
\end{table}


We categorize the manually fixed errors into five distinct groups, as summarized in Table~\ref{tbl-error}. 

First, the Type Error (TE) dominates all error types in math and ramfs. These errors include type mismatches, undefined types, or types that fail to implement required traits and are common in Rust due to the language's strong typing and numerous type-related constraints. For instance, Figure~\ref{fig-te} shows that any type mismatch results in a compile-time error. 

Second, Missing Imports (MI) are the second most frequent error. It includes issues such as missing trait imports, traits not in scope, or methods unavailable due to unimplemented traits. This error occurs because our translation process involves resolving many dependencies, and while most have been addressed using two dependency probers (Section~\ref{sec:design}), some remain unresolved. 

Third, Changed Symbol (CS) errors account for a moderate proportion of errors, reflecting the ``hallucination'' phenomenon commonly observed in LLM outputs. For example, Figure~\ref{fig-cs} illustrates that the LLM incorrectly identifies the type name as Inode instead of inode. 

Fourth, we discover a semantic mistake in the expression $size \ \&\  -size$ because it behaves differently in C than in Rust. As shown in Figure~\ref{fig-sem}, the LLM does not account for Rust’s wrapping rules and generates incorrect code. In Rust, writing $-size$ when $size$ is unsigned can lead to unexpected results. The correct expression is $size \ \& \  !size.wrapping\_add(1)$, which ensures the intended bitwise operation without relying on negative values in unsigned arithmetic.

Table~\ref{tbl-error} also include an ``Other'' column that includes the use of unsafe code, miscellaneous syntax errors, such as missing modules, crates, or other structural elements essential for the correct compilation and execution of the translated Rust code.



\subsection{RQ3: Comparison between function-level translation and whole-module translation}

\vspace{2mm}
Table~\ref{tab:function-level} presents the results of function-level translation for three datasets, compared to module-level translation using GPT-4o in a single pass. We have three key observations. First, function-level translation \system\ significantly reduces the laziness problem entirely, achieving a laziness rate of zero for sort and ramfs. This demonstrates its effectiveness in addressing incomplete or inaccurate translations often seen in whole-module translation. Second, Function-level translation yields higher CodeBLEU scores and requires fewer manually modified lines (MMLs). This can be attributed to the inherent difficulty for LLMs to accurately translate entire modules in a single attempt, as the probability of all tokens being correct is low. By translating on a function-by-function basis, \system\ allows for human verification and intervention at each step, significantly increasing the likelihood of producing accurate and high-quality code.

\subsection{RQ4: The effectiveness of the \textit{Repair} component} 
\vspace{2mm}

Table~\ref{tab-new-repair} highlights the impact of the iterative repair process in addressing LLM limitations when translating complex functions. We observe two key outcomes. First, \textit{Repair} is effective in fixing errors that the LLM fails to resolve during translation. After three repair iterations, the \textit{Repair} component significantly increases the CSR from 7/18 to 15/18 before reaching a plateau. Second, \textit{Repair} successfully resolves specific error types that LLMs cannot address, such as symbol misuse involving variable names (E0425), type names (E0412), crates/modules (E0433), and fields (E0609). However, as discussed in RQ2, certain errors still require manual intervention to ensure correctness.

\section{Related Work}
\label{sec:related}

Code translation is a long-discussed topic and is highly beneficial to software development, particularly in enabling cross-platform compatibility, code reuse, and modernization. To achieve effective code translation, there are two main categories of approaches: rule-based approaches and LLM-based approaches.


\paragraph{Rule-based approaches.} Tools like transpilers and source-to-source compilers employ program analysis techniques to convert code between programming languages. These tools typically involve detailed syntactic and semantic analysis of the source code to identify equivalent constructs in the target language. For example, C2Rust~\cite{c2rust} and CxGo~\cite{CxGo} are well-known transpilers for translating C programs to Rust and Go, respectively. Similarly, Sharpen~\cite{Sharpen} and Java2CSharp~\cite{Java2CSharp} are designed to convert Java code to C\#. 

Rule-based approaches maintain close alignment with the structure and semantics of the original code. In addition, thanks to program analysis and compiler-based tools, traditional approaches often produce reliable and consistent output that is crucial for system-level translations, and the translations are predictable and deterministic, providing clearer debugging and traceability. But traditional approaches may struggle with advanced language features or idioms that don’t have straightforward equivalents (e.g., pointer arithmetic in C to safe Rust references). Besides, traditional approaches lack idiomatic translation, which can reduce maintainability. Furthermore, building effective transpilers is labor-intensive, as each language feature and idiom must be carefully mapped, and any new language versions require updates.

\paragraph{LLM-based approaches.} The emergence of LLM has introduced an alternative paradigm for code translation by leveraging advances in natural language processing and machine learning, allowing for flexible, data-driven solutions that improve over time. Early approaches included lexical statistical machine translation~\cite{DBLP:conf/oopsla/KaraivanovRV14, DBLP:conf/sigsoft/NguyenNN13} and tree-based neural networks~\cite{DBLP:conf/nips/ChenLS18, DBLP:journals/corr/abs-1807-01784} for program language translation. Other techniques leverage deep learning and unsupervised learning~\cite{DBLP:conf/nips/RoziereLCL20} to translate languages like C++, Java, and Python. More recently, large language models (LLMs) have become prominent, with models such as StarCoder~\cite{DBLP:journals/tmlr/LiAZMKMMALCLZZW23}, PolyCoder~\cite{DBLP:conf/pldi/Xu0NH22}, SantaCoder~\cite{DBLP:journals/corr/abs-2301-03988}, CodeGen~\cite{DBLP:journals/ijseke/LiKZWLLX23}, BLOOM~\cite{DBLP:journals/corr/abs-2211-05100}, CodeT5~\cite{DBLP:conf/emnlp/0034WJH21}, CodeX~\cite{DBLP:journals/corr/abs-2107-03374}, GPT-4~\cite{DBLP:journals/corr/abs-2303-08774}, and Llama 3~\cite{DBLP:journals/corr/abs-2407-21783} applied to tasks such as code synthesis, completion, and translation. For example, CodeGeeX~\cite{DBLP:conf/kdd/ZhengXZDWXSW0LS23} has shown the capability to translate between languages, generating idiomatic code across languages like Rust, C, and more.

LLM-based approaches can learn to translate diverse language features and idioms, generating idiomatic code in the target language and they can automatically adapt to updates. However, due to the data-driven nature of LLMs, translations may occasionally produce syntactically or semantically incorrect code, especially in complex or ambiguous cases. Besides, deep learning models are often treated as ``black boxes'', making it difficult to trace or modify specific aspects of their translations.

Both rule-based and deep-learning-based approaches play vital roles in advancing code translation, each suited to different translation needs and complexity levels. Therefore, in this paper, we propose to combine these two approaches and get the best of both worlds.
\section{Discussion}

\vspace{2mm}
\label{sec:discussion}

In this section, we examine the limitations of \system\ and propose potential directions for future work.

\subsection{Limitations}

Despite the promising results achieved with \system, several limitations remain.

One primary limitation of \system\ lies in its generalizability. While \system\ successfully translates a variety of C functions into functional Rust code, our experiments focus on a narrow subset of Linux kernel modules, namely, math, sort, and ramfs. However, we carefully chose these modules to represent diverse functionality and complexity, allowing us to evaluate \system\ across different translation scenarios.

The efficiency of \system\ is another limitation, particularly in terms of time complexity, which can be variable and occasionally unbounded without meticulous process management. First, the time required for automated components such as parsing, translation, and testing is long and varies significantly. For instance, the iterative processes in components such as the \textit{C Dependency Prober}, \textit{Rust Dependency Prober}, and \textit{Repair} can extend the processing time. However, we find that rule-based iterations, such as external symbol probing, typically complete within 15–20 iterations (each lasting a few seconds). For LLM-based iterations, we limit the iteration time to constrain the overhead such as three times in program repair. Second, the time required for human effort does not always align with the Manually Modified Lines (MMLs). For example, addressing 100 MMLs might take only a few minutes, while fixing five complex MMLs could require days of debugging and repair. Fortunately, we address extreme cases of human labor by leveraging unsafe Rust generated by tools like C2Rust, bypassing prolonged debugging efforts when necessary.

\subsection{Future work}

Several promising directions exist for enhancing \system\ in future work. First, the integration of LLMs specifically trained on low-level systems code could improve the quality and accuracy of Rust code generation. By incorporating models specifically fine-tuned for kernel-level C code, we could potentially reduce the need for manual intervention and improve the overall reliability of the system. Second, automating the program repair process more effectively is crucial for reducing the reliance on human input. This could be achieved through the development of more advanced static analysis tools and program synthesis techniques that could detect and resolve errors automatically, even for complex kernel-specific issues. Third, to comprehensively evaluate and refine \system, we plan to systematically translate additional modules from other codebases.
\section{Conclusion}
\label{sec-conclusion}
In this paper, we have introduced \system, a novel LLM-based C-to-Rust translation tool that addresses the ``laziness'' issue commonly observed with large language models when processing large code modules. By decomposing the translation task into function-level units and using a combination of rule-based and LLM-based strategies, \system\ not only maintains high accuracy but also enhances the generation of safe Rust code, which is essential for the reliability of system-level software. Our evaluation of \system\ has demonstrated its effectiveness through significant reductions in manual intervention, decreasing the need for human corrections to less than 15\% of the translated code lines. Moreover, \system\ produced a high proportion of safe Rust code, improving memory safety over existing approaches.

\bibliographystyle{ACM-Reference-Format}
\bibliography{references}



\end{document}